\documentclass[pre,twocolumn,aps,showpacs,amsmath,floatfix]{revtex4}
\usepackage{graphicx}
\usepackage{dcolumn}
\usepackage{amsmath}
\begin{document}
\newcommand{\be}{\begin{equation}}
\newcommand{\ee}{\end{equation}}

\title{Bulk and surface magnetoinductive breathers in binary metamaterials}

\author{M. I. Molina$^{1}$, N. Lazarides$^{2,3}$, and G. P. Tsironis$^{2}$}

\affiliation{$^{1}$Departamento de F\'{\i}sica, Facultad de
Ciencias, Universidad de Chile, Casilla 653, Santiago, Chile\\
$^{2}$Department of Physics, University of Crete and Institute 
of Electronic Structure and Laser, 
Foundation for Research and Technology-Hellas, P.O. Box 2208, 71003 Heraklion, Greece\\
$^{3}$Department of Electrical Engineering, Technological Educational 
Institute of Crete, P.O. Box 140, Stavromenos, 71500 Heraklion, Crete, Greece}
\date{\today}

\begin{abstract}
We study theoretically the existence of bulk and surface discrete
breathers in a one-dimensional magnetic metamaterial comprised  
of a periodic binary array of split-ring resonators. 
The two types of resonators differ in the size of their slits and this 
leads to different resonant frequencies.  
In the framework of the rotating-wave approximation (RWA)
we construct several types of breather excitations for both the energy-conserved
and the dissipative-driven systems by continuation of trivial 
breather solutions from the anticontinuous limit to finite couplings.
Numerically-exact computations that integrate the full model equations
confirm the quality of the RWA results.
Moreover, it is demonstrated that discrete breathers can
spontaneously appear in the dissipative-driven system as a results 
of a fundamental instability.
\end{abstract}


\maketitle
\section{Introduction}
Discrete breathers (DBs) or intrinsic localized modes, 
are time-periodic and spatially localized excitations that may be 
produced generically in discrete arrays of weakly coupled nonlinear 
elements \cite{ST,Flach}. A large body of theoretical work has produced means of precise
numerical analysis of coupled oscillator systems with breathers
both in the Hamiltonian as well as dissipative 
regimes~\cite{MacKay,Aubry1997,Marin1996,Marin2001,Zueco}.
Breathers  have been experimentally observed in several diverse systems, 
including optical waveguides systems \cite{Eisenberg}, 
solid-state systems \cite{Swanson,Russell}, antiferromagnetic chains \cite{Schwarz}, 
Josephson junction arrays \cite{Trias}, 
and micromechanical oscillators \cite{Sato2003,Sato2007},  among others.
Discrete breathers  can be generated spontaneously in a lattice either through stochastic
mechanisms \cite{TA,R} or
by purely deterministic mechanisms \cite{Dauxois,Sato2006,Hennig2007,Hennig2008} in a 
process by which energy, initially evenly distributed in a nonlinear lattice,
can be localized into large amplitude nonlinear excitations.
Indeed, it has been demonstrated experimentally \cite{Sato2003,Sato2006} that 
the standard modulational instability (MI) mechanism in dissipative systems
driven by an alternating term can initiate that process by the formation
of low-amplitude breathers. The energy exchange between those low-amplitude DBs
favors the higher-amplitude ones, resulting eventually in the formation
of a few high-amplitude DBs.

A few years ago a whole new class of artificially structured
materials, referred to as metamaterials, was discovered; the latter are  comprised of discrete elements
and exhibit electromagnetic properties not available in naturally occurring materials.
A subclass of those metamaterials, the magnetic metamaterials (MMs),
exhibit significant magnetic properties
and sometimes even negative magnetic response up to Terahertz (THz) and 
optical frequencies \cite{Gorkunov2002,Linden}. 
The most common realization of a MM is composed of 
periodically-arranged electrically small sub-wavelength particles, referred
to as split-ring resonators (SRRs) \cite{Yen,Katsarakis}. In its simplest form,
each of those resonators is just a highly conducting metallic ring with one slit.
The MM thus built can become nonlinear either by the insertion of a nonlinear 
dielectric \cite{Hand} or a nonlinear electronic component (e.g., a varactor diode)
\cite{Powell,Shadrivov2006} in the slit of each SRR, resulting in 
voltage-dependent SRR capacitance. In  microwave frequencies, such a MM has been 
realized \cite{Shadrivov2008} and is dynamically
tunable  by varying the input power.
The combination of nonlinearity and the discreteness that is inherent in those
metamaterials, makes possible the generation of nonlinear excitations in the 
form of DBs.
The existence and stability of DBs in nonlinear SRR-based MM models,
that are localized either in the bulk \cite{Lazarides2006,Eleftheriou2008}
or at the surface  \cite{Lazarides2008,Eleftheriou2009} of the MM,
have been demonstrated numerically.
Moreover, domain walls \cite{Shadrivov2006b}
and envelope solitons \cite{Kourakis} may also be excited in such systems.
The surface DBs in MMs are very similar with the surface modes observed in
discrete waveguide arrays \cite{Molina2006b,Kivshar}.

Recently, a novel MM comprised of two types of SRRs was investigated 
theoretically \cite{Gorkunov2006}, and it was demonstrated that in the nonlinear 
regime such binary MMs are perfectly suited for the observation of phase-matched
parametric interaction and enhanced second harmonic generation (SHG).
In the present work we extend the previous studies on DB generation in model 
SRR-based MMs to the case of a one-dimensional (1D) binary MM with on-site
cubic nonlinearity.
In practice, a binary MM can be constructed in many different ways,
by changing for example one or more of the material and/or the geometrical  
parameters of the SRRs belonging to one type (say type $a$), 
with respect to the same parameters of the SRRs belonging to the other type
(say type $b$). Here we allow for different slit-widths for the two types 
of SRRs, which makes their resonance frequencies differ by a factor
that we call resonance frequency mismatch (RFM). The considered binary MM is formed 
by type $a$ SRRs at the even-numbered sites and type $b$ SRRs at the odd-numbered 
sites of a periodic 1D array.
In the next section we give  the model equations that describe the dynamics
of the binary MM and we obtain the corresponding linear dispersion relation
for magnetoinductive waves in that medium \cite{Shamonina,Shadrivov2007}.
In Sections III and IV we construct, using the rotating wave approximation (RWA),
several types of Hamiltonian and dissipative breathers (DDBs), respectively.
The dynamic stability of those DBs is discussed in Section V 
where the full model equations are integrated numerically.
In most of the investigated cases the numerics confirm the quality of the RWA results. 
Moreover, in that Section we also demonstrate the spontaneous generation of DDBs
induced by MI, using frequency chirping of the driving field.
That procedure has been used in actual experiments for the generation of 
high-amplitude DDBs in micromechanical cantilever oscillator arrays \cite{Sato2003},
and perhaps it could be also used in similar experiments involving nonlinear 
binary MMs.
We finish in section VI with the conclusions.
\begin{figure}[t!]
\noindent
\includegraphics[scale=0.3,angle=0]{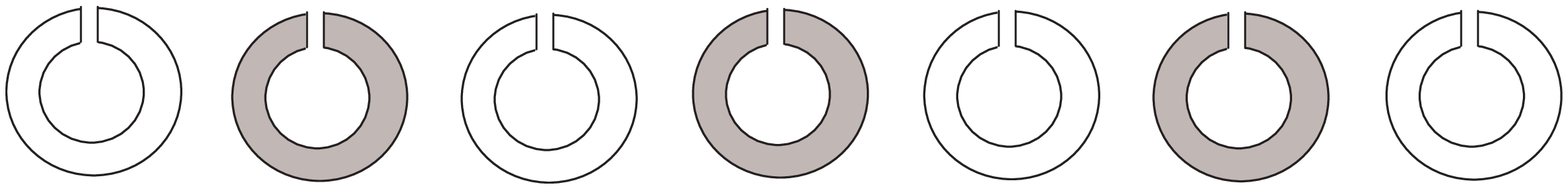}
\caption{One-dimensional binary array of split-ring resonators.}
\label{fig1}
\end{figure}

\section{Model Binary Metamaterial and Linear Dispersion
}
Consider a 1D SRR-based MM comprised of nonlinear units shown 
schematically in Fig. 1.
Each nonlinear SRR in the array can be mapped to a nonlinear 
resistor-inductor-capacitor (RLC) circuit featuring self-inductance $L$,
ohmic resistance $R$, and nonlinear (voltage-dependent) capacitance 
$C(|{\bf E}|^2) \propto \epsilon (|{\bf E}_g|^2)$, where $\epsilon$
is the field-dependent permittivity of the infilling dielectric,
${\bf E}$ is the electric field, and ${\bf E}_g$ is the electric field
induced along the SRR slit. We assume that the latter originates from
an alternating magnetic field that is applied to the MM perpendicularly 
to the SRR planes, and it is proportional to the voltage $U$ across the 
slit.

Let us for the moment ignore the nonlinearity and set $C=C_l$, with 
$C_{l}$ is the linear capacitance that is built up across the slit.
Just like an RLC resonator, the SRRs exhibit an inductive-capacitive
resonance at frequency $\omega_R  \simeq 1/\sqrt{L\ C_l}$ 
(for $R\simeq 0$, implying low Ohmic losses). 
For a circular SRR with rectangular cross-section, the parameters 
$L$, $R$, and $C_{l}$ of the equivalent RLC circuit can be estimated 
from the relations \cite{Gorkunov2002}
\begin{eqnarray}
\label{1}
L = \mu_{0} r \left[\log\left( \frac{16 r}{h} \right)-1.75 \right], \, \, 
C_l =\epsilon_0 \epsilon_l \frac{\pi h^2}{4 d}, \, \,  
R = {8 \rho r \over{h^{2}}}
\end{eqnarray}
where $\epsilon_0$ and $\mu_0$ are the permittivity and permeability
in vacuum, respectively,
$\epsilon_l$ is the linear relative dielectric permittivity of the infilling
dielectric, $r$ is the average  SRR radius, $h$ is the diameter of the metal wire,
$d$ is the slit width of the SRR, and $\rho$ the (material-dependent)
SRR resistivity.
Neighboring SRRs in an array are magnetically coupled through their
mutual inductance $M$, that is approximately given by
\begin{eqnarray}
\label{2}
M \approx {\mu_{0} \pi r_a^{2} r_b^{2}\over{4 D^{3}}}
\end{eqnarray}
where $D$ is their center-to-center distance, and $r_a, r_b$ are their average
radii.
In order to construct a binary array, we have to change one or more of the 
material and/or geometrical parameters of the SRRs that are going to be of one
type, with respect to the same parameters of the SRRs that are going to be of
the other type.
As it can be observed from Eqs. (\ref{1}) and (\ref{2}) \newline
 $\bullet$ A change in the SRR radius $r$ affects $L,M$ and $R$.  \newline
 $\bullet$ A change in $h$ affects $L, C_{l}$ and $R$.  \newline
 $\bullet$ A change in $\epsilon_l$ affects $C_l$ which in turn, 
     implies a change in the nonlinear response.  \newline
 $\bullet$ A change in $d$ affects $C_l$ and slightly $R$ and $L$.  \newline
 $\bullet$ A change in resistivity $\rho$ affects $R$. 

Obviously there many possibilities for constructing two types of SRRs and consequently
a binary array. Here we make the relatively simple choice to create two types 
of SRRs by considering different slit-widths, i.e., $d_a$ for type $a$ and $d_b$
for type $b$.
Thus, the linear capacitances of type $a$ and $b$ SRRs become respectively 
$C_a$ and $C_b$, resulting in different resonance frequencies
$\omega_a=1/\sqrt{L C_a}$ and $\omega_b=1/\sqrt{L C_b}$. 

Now let us return to the nonlinear problem, and assume that the slits of all the 
SRRs in the array are filled with a Kerr-type dielectric.
Then, the charge $Q_n$ accumulated in the capacitor of the $n$th SRR is \cite{Zharov}
\begin{equation}
\label{3}
Q_n = C_n \left(1 + \chi {U_n^2\over{U_c^2}}\right) U_n
\end{equation}
where $C_n =C_a$ ($C_n =C_b$) for SRRs at even- (odd-) numbered sites of the array,
and $\chi = \alpha / (3 \epsilon_l)$ is the dimensionless nonlinearity coefficient,
with $\alpha=+1$ ($\alpha=-1$) for a self-focusing (self-defocusing) dielectric.
The above equation leads to the following approximate form for the voltage $U_n$ 
across the slit of the $n$th SRR
\begin{equation}
\label{4}
 U_n \approx {Q_n\over{C_n}}
 \left[ 1 - \chi {1\over{U_c^2}}\left({Q_n\over{C_n}} \right)^{2} \right]
\end{equation}
where $U_{c}$ is a characteristic (large) voltage.
Then, the coupled equations describing the charge dynamics in the nonlinear MM
that is placed in an alternating magnetic field are generally written as
\begin{eqnarray}
\label{5}
  \frac{d^2}{dt^2}\left[ M  Q_{2n-1} +L_{2n} Q_{2n} +M Q_{2n+1} \right]
  +R_{2n} \frac{d}{dt} Q_{2n}  \nonumber \\ 
  +\frac{1}{C_{2n}} Q_{2n} 
  -\chi {1\over{U_{c}^{2}}}\left({Q_{2n}\over{C_{2n}}}\right)^{3} =F(t)  \\
\label{6}
   \frac{d^2}{dt^2}\left[  M Q_{2n} +L_{2n+1} Q_{2n+1} +M Q_{2n+2} \right]
   +R_{2n+1}  \frac{d}{dt} Q_{2n+1} \nonumber \\ 
   \frac{1}{C_{2n+1}} Q_{2n+1}
   -\chi {1\over{U_{c}^{2}}}\left({Q_{2n+1}\over{C_{2n+1}}}\right)^{3} =F(t) , 
\end{eqnarray}
where we assume that $R_m =R$, $L_m =L$, and 
\begin{equation}
  \label{6.1}
  F(t) = {\cal E}_{0} \sin(\omega t) .
\end{equation}  
The above equation gives the electromotive force (emf) of amplitude ${\cal E}_{0}$
and frequency $\omega$ that is excited in each SRR due to the action of the field.
Let us define the quantities
\begin{equation}
\label{7}
  Q_c=\sqrt{C_a C_b} U_c,  \ \ \ \tau=\sqrt{\omega_a \omega_b} t\equiv \omega_{0} t, 
   \ \ \ \gamma=R \sqrt{ \frac{\sqrt{C_a C_b}}{L} } .
\end{equation}
With the above definitions, Eqs. (\ref{5}) and (\ref{6}) can be written in normalized
form as
\begin{eqnarray}
   \label{8}
    {d^{2}\over{d \tau^2}}[\lambda q_{2n-1} + q_{2n}+\lambda q_{2n+1} ] 
    +\delta q_{2n} -\chi \delta^3 q_{2n}^{3} =   \nonumber \\
    -\gamma {d q_{2n}\over{d \tau}}
    +\epsilon_0 \sin(\Omega \tau) , \\
    \label{9}
    {d^{2}\over{d \tau^2}}[\lambda q_{2n+1} + q_n+\lambda q_{2n+2} ] 
    +\frac{q_{2n+1}}{\delta} 
    -\chi \frac{q_{2n+1}^3}{\delta^3} =  \nonumber \\ 
    -\gamma {d q_{2n+1}\over{d \tau}}
    +\epsilon_{0} \sin(\Omega \tau) , 
\end{eqnarray}
where $\lambda \equiv M/L$ is the dimensionless coupling parameter, 
$\Omega =\omega/\omega_{0}$ is the dimensionless driving frequency,
$\delta \equiv \omega_{a}/\omega_{b}$ is the RFM ratio,
$\epsilon_0={\cal E}_{0}/U_c$, and $n$ is an integer.
From Eqs. (\ref{8}) and  (\ref{9}) we can see that
the change in the linear capacitances also affects the nonlinear terms, 
and that, actually, the latter are affected much more than the linear ones. 
The changes that are caused to the terms proportional to $R$ and $L$ 
are of higher order and thus they are neglected.
As the RFM $\delta$ increases, the resonance frequency as well as the nonlinear 
term of the even-numbered site SRRs increase,
while at the same time the resonance frequency and the nonlinear term 
of the odd-numbered site SRRs decrease.
The inductive coupling parameter $\lambda$ can be either positive or negative
depending on whether the array geometry is of
the ``axial'' or ``planar'' type \cite{Eleftheriou2008}.

Such MMs and other systems with magnetically coupled elements, 
support in the low-amplitude (linear) limit a new kind of waves, 
the magnetoinductive waves \cite{Shamonina,Shadrivov2007}.
\cite{Shamonina,Shadrivov2007}.
In the present case, the frequency dispersion for linear magnetoinductive 
waves is obtained by substituting
\begin{eqnarray}
 \label{10}
   q_{2n} = a\, e^{i(2n\kappa -\Omega \tau)}, \qquad 
   q_{2n+1} = b\, e^{i((2n+1)\kappa -\Omega \tau)} ,
\end{eqnarray}
where $\kappa$ is the normalized wavenumber,
into the linearized Eqs. (\ref{8}) and (\ref{9}),
without the dissipative and the external driving terms
($\gamma=0$ and $\epsilon_0=0$)
\begin{equation}
 \label{11}
  \Omega_{\pm}^{2} = {{[\delta+(1/\delta)] 
  \pm \sqrt{[ \delta+(1/\delta)]^2 +4(1 - 4 \lambda^2 \cos^2 \kappa)}}
              \over{2 (1 - 4 \lambda^2 \cos^2 \kappa)}}
\end{equation}
\begin{figure}[h!]
\noindent
\includegraphics[scale=.6,angle=0]{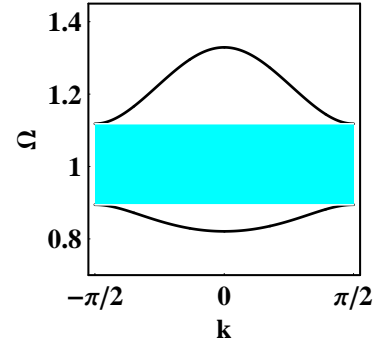}
\caption{
(Color online)
Linear dispersion relation for $\delta=0.8$ and $\lambda=-0.2$.}
\label{fig2}
\end{figure}
The dispersion curves for a particular choice of RFM
$\delta$ and 
coupling parameter $\lambda$ are shown in Fig. 2. 
Even though we can think of each linear SRR as an RLC circuit,
i.e., an electromagnetic  oscillator, we notice that the dispersion curves 
for the coupled array do not contain any acoustic-like branch; 
rather both curves are of the ``optical'' type: $\lim_{k\rightarrow 0} w(k)\neq 0$. 
This is due to the particular (inductive) nature of the coupling between the SRRs. 
There are now three regions where we can look for breathers in the nonlinear case: 
Above and below the bands, and in the Bragg gap (BG) in between. 
In the absence of RFM ($\delta=1$) we recover the single band 
for the `monoatomic' MM \cite{Eleftheriou2008}. 
As $\delta$ diverges from unity the BG increases, pushing the $\Omega_{+}$ 
branch upwards and the $\Omega_{-}$ branch downwards.

Eqs. (\ref{8}) and  (\ref{9}) can be written conveniently in the compact form
\begin{eqnarray}
 \label{12}
     {d^{2}\over{d \tau^2}}[\lambda q_{n-1} + q_n+\lambda q_{n+1} ] 
       +\omega_n^2 q_n -\chi \omega_n^6 q_n^{3} = \nonumber \\
    -\gamma {d q_n\over{d \tau}}
   +\epsilon_{0} \sin(\Omega \tau) , 
\end{eqnarray}
where $\omega_n^2 = \delta$  ($\omega_n^2 = 1/\delta$) for even (odd) $n$.
Without dissipation and external driving, the earlier 
equation can be obtained from the Hamiltonian ${\cal H} = \sum_{n} {\cal H}_n$,
where the discrete Hamiltonian density ${\cal H}_n$ is given by 
\begin{eqnarray}
 \label{17} 
   {\cal H}_n = \frac{1}{2} \left\{
     \dot{q}_{n}^2
    +\lambda \, \dot{q}_{n} ( \dot{q}_{n-1} + \dot{q}_{n+1} )
   \right\} +  V_{n} . 
\end{eqnarray}		       
The last term on the right-hand side of the earlier equation is 
the nonlinear on-site potential which is given by
\begin{eqnarray}
 \label{15}
  V_{n} \equiv  V ( q_n ) =\frac{1}{2} (\omega_n q_n)^2
   \left[ 1 -\frac{1}{2} \chi \omega_n^2 (\omega_n q_n)^2 \right] .
\end{eqnarray} 
The Hamiltonian ${\cal H}$ is actually the conserved energy of the 
lossless system in the absence of any driving terms system. 
For the dissipative system, ${\cal H}$ is also useful since 
its time-average per period gives correctly the average energy
per period for that system.

\section{Hamiltonian Breathers in the Rotating-Wave approximation
}
A standard method of DB construction in Hamiltonian systems,
that gives numerically exact results up to arbitrary precision, 
uses the Newton's  method \cite{Marin1996, Aubry1997}, which has been 
applied successfully for DB generation in MMs \cite{Lazarides2006,Eleftheriou2008}.
In this Section, we use the RWA method that keeps a simple physical picture
and moreover can produce quite accurate results.
According to the simplest version of that method, one looks for stationary 
solutions of the system that are separable with an assumed time-dependence 
(e.g., sinusoidal) of the form $q_n (\tau) =q_n\, \sin(\Omega \tau)$.
Direct substitution of $q_n (\tau)$ into Eq. (\ref{12}) with the approximation
$\sin(x)^3 \approx (3/4) \sin(x)$ gives an algebraic system of nonlinear equations 
for the $q_n$s that reads
\begin{eqnarray}
\label{18}
   -\Omega^2 ( \lambda q_{n+1} + q_n + \lambda q_{n-1} ) + \omega_n^2 q_n 
   -(3/4) \chi \omega_n^6 q_n^3 = 0 .
\end{eqnarray}
In the anticontinuous limit ($\lambda \rightarrow 0$) the earlier equation 
has the solutions
\begin{equation}
\label{19}
    q_{n}=0 \qquad \mbox{or} \qquad  
    q_{n}^{2}={\omega_n^2-\Omega^2 \over{(3/4) \chi \omega_n^6}} .
\end{equation}
According to Eqs. (\ref{19}), 
we have the following interesting possible scenarios:

(i) $\alpha>0$. Then, $q_{n}^{2}>0$ for all $n$, 
  if $\Omega^2<\mbox{Min}\{\delta,1/\delta\}$.

(ii) $\alpha<0$. 
  Then $q_{n}^{2}>0$ for all $n$ provided 
  $\Omega^2 > \mbox{Max}\{ \delta, 1/\delta\}$.

(iii) In the intermediate case where 
   $\mbox{Min}\{\delta,1/\delta\}<\Omega^2<\mbox{Max}\{\delta,1/\delta\}$, 
   what happens is that $q_{2n}>0$ and $q_{2n+1}=0$, or the converse, 
   depending upon the sign of $\alpha$.

The RWA method can be used for the construction of DBs both on the 
`surface' and the bulk of the energy-conserved  binary MM. 
For a 1D MM, a surface localized DB obviously corresponds to an 
edge state, i.e., a state with maximum amplitude at either of the two
ends of the array. A bulk DB, on the other hand, is meant to be a DB 
whose maximum amplitude is far from the end-points of the array.
However, the procedure of obtaining DBs by the RWA method,
both at the surface or in the bulk, proceeds in 
the same way.
The first step is to set up a trivial DB. We first choose its central site, 
i.e., the site
where the DB shall have its maximum amplitude. Suppose that the central site is
taken at $n=n_B$ where the `coordinate' $q_n =q_{n_B}$ and set all 
the $q_n$ for $n\ne n_B$ equal to zero. 
The value of $q_{n_B}$ is calculated from Eq. (\ref{19}),
with an appropriately chosen $\Omega$.
That solution is subsequently continued for finite couplings 
up to a maximum value $\lambda=\lambda_{max}$ where DBs cease to exist.
Usually we consider a DB to be localized around the site where it exhibits 
its maximum amplitude.
For a finite array, the boundary conditions that should
be imposed to the dynamic equations resulting from the Hamiltonian (\ref{17}) should be specified.
For DBs excited in  the bulk one may use either periodic or open-ended
boundary conditions, since DBs are highly localized entities and are not
affected by the boundaries. However, for surface DBs the termination of the 
structure should be taken into account, and for that purpose we should use
open-ended boundary conditions. In the following, we always use that type
of boundary conditions, i.e., $ \ q_0 =0, \ \ \ q_{N+1}=0$, 
where $N$ is the total number of SRRs in the binary array.
\begin{figure}[t!]
\begin{center}
\includegraphics[scale=0.35,angle=0]{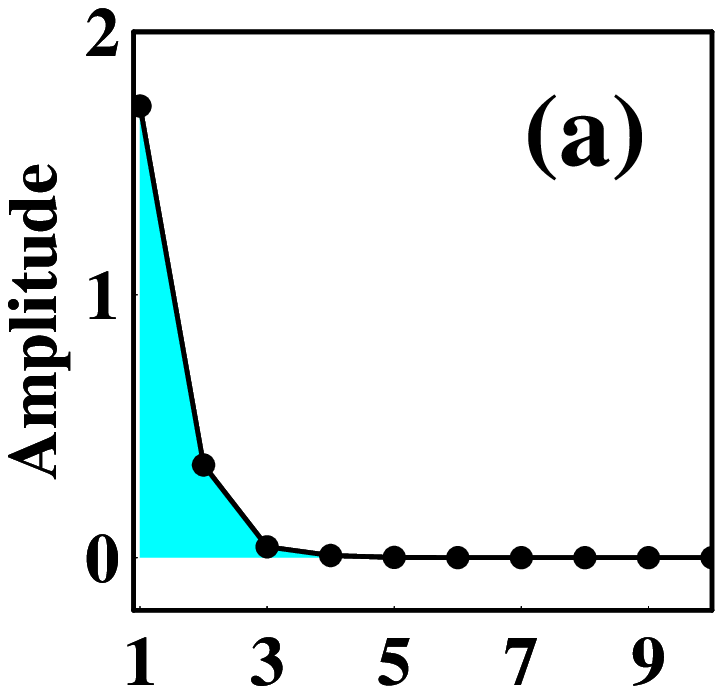}
\includegraphics[scale=0.35,angle=0]{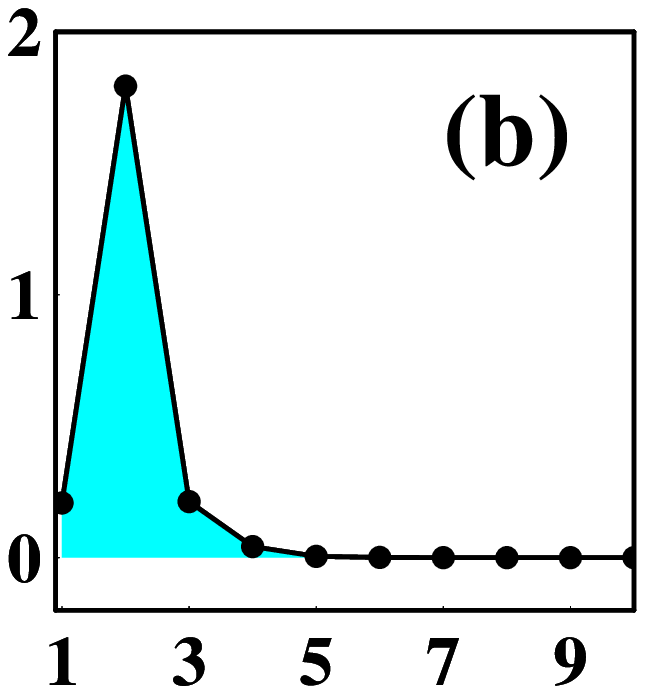}
\includegraphics[scale=0.35,angle=0]{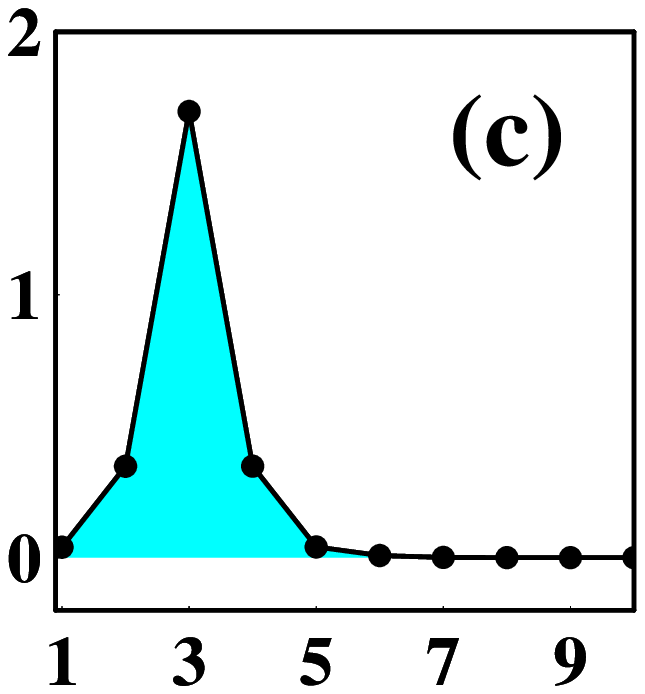}
\includegraphics[scale=0.35,angle=0]{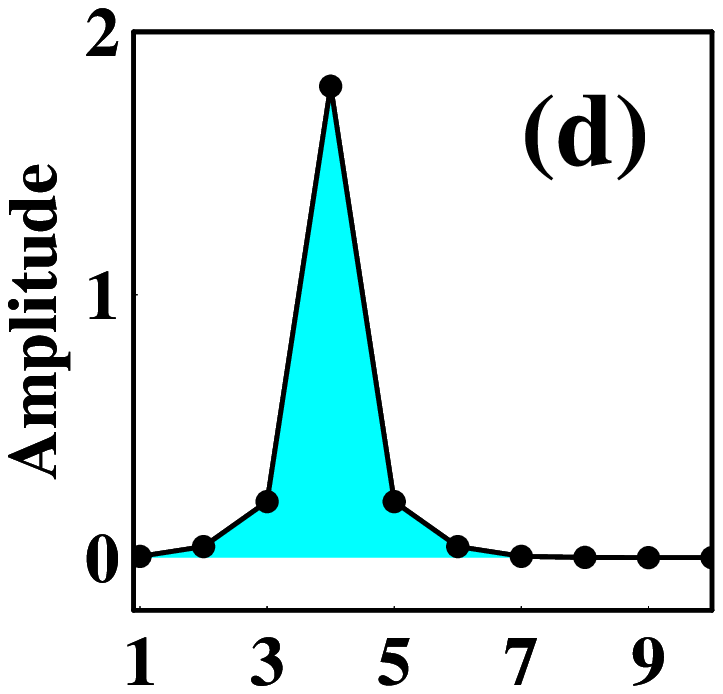}
\end{center}
\caption{
(Color online)
Typical Hamiltonian surface breather profiles in a magnetoinductive binary chain for 
$\delta=0.9$, $\lambda=0.1$, $\Omega=0.77$, $\chi=1/6$, obtained by the RWA.
}
\end{figure}

{\em Surface breathers.-} 
Typical surface-localized, single-site DBs, 
that are generally very similar to the ones examined for
a discrete nonlinear Schr\"odinger (DNLS) model for a  semi-infinite binary 
waveguide array \cite{Molina2006}, are displayed in Fig. 3.
The unstaggered modes shown there originate in the lower gap region
$0<\Omega<{\mbox Min}[\delta,1/\delta]$. There are also staggered modes (not
shown) originating from the Bragg gap region that constitute
magnetoinductive Tamm states
\cite{Lazarides2008,Eleftheriou2009}.
From the surface DBs shown in Fig. 3, only one of them (Fig. 3a)
corresponds to a truly surface state, since it is localized
exactly at the left end of the array ($n=1$). 
The next two (Figs. 3b and 3c) can also be  characterized as surface DBs,
since they are localized
very close to the surface ($n=2$ and $n=3$, respectively),
but actually they
are cross-over states between surface and bulk DBs.
Since the DBs shown here are highly localized, they obtain
their bulk form within a distance of only a few sites from the surface,
so that the DB shown in Fig. 3d (localized at $n=4$) can be
considered as a bulk DB.
With the appropriate choice of the initial conditions we may also 
construct multi-site surface DBs such as those shown in Fig. 4, that 
remind four-site antisymmetric DB excitations. Of course, close to the
surface, we cannot obtain exact antisymmetric DBs.
\begin{figure}[t!]
\begin{center}
\includegraphics[scale=0.4,angle=0]{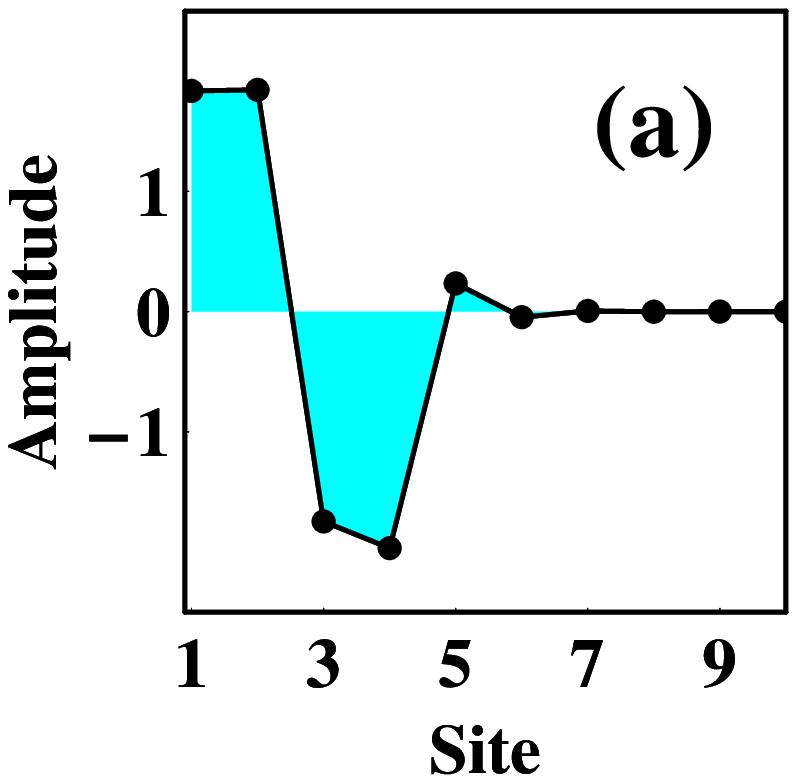}
\includegraphics[scale=0.4,angle=0]{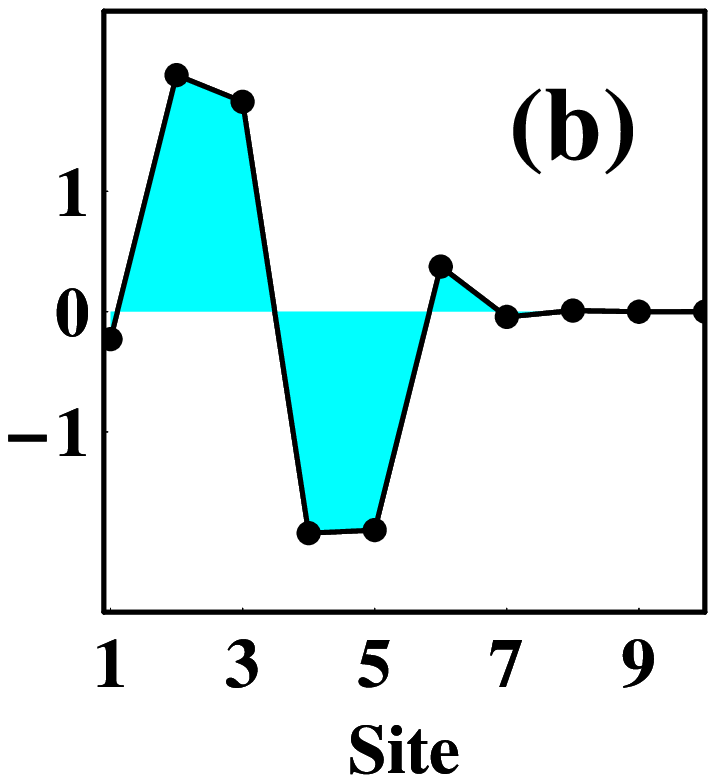}
\includegraphics[scale=0.4,angle=0]{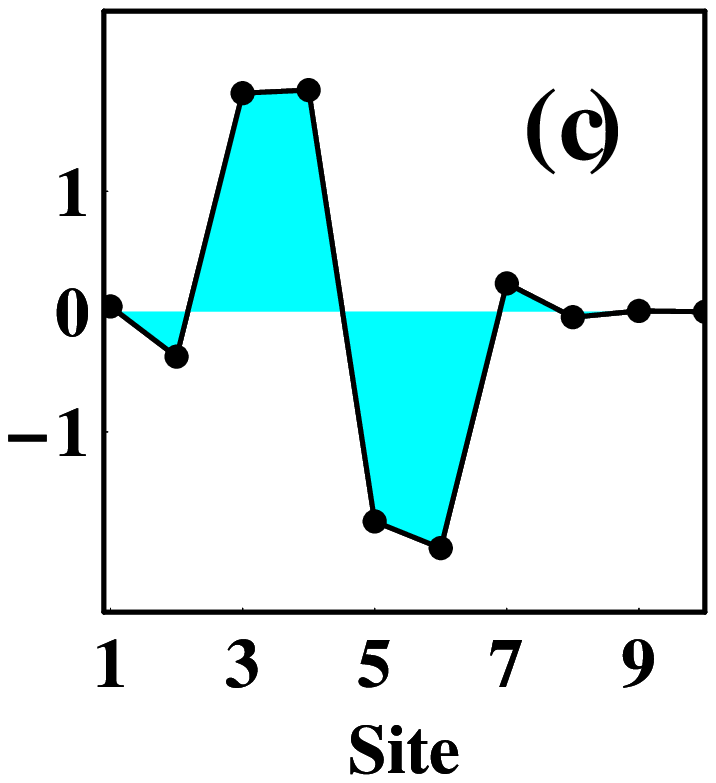}
\end{center}
\caption{
(Color online)
Typical Hamiltonian surface multi-site breather profiles in a magnetoinductive 
binary chain for 
$\delta=0.9$, $\lambda=-0.1$, $\Omega=0.77$, $\chi=1/6$, obtained by the RWA.
}
\end{figure}
\begin{figure}[t]
\begin{center}
\includegraphics[scale=0.4,angle=0]{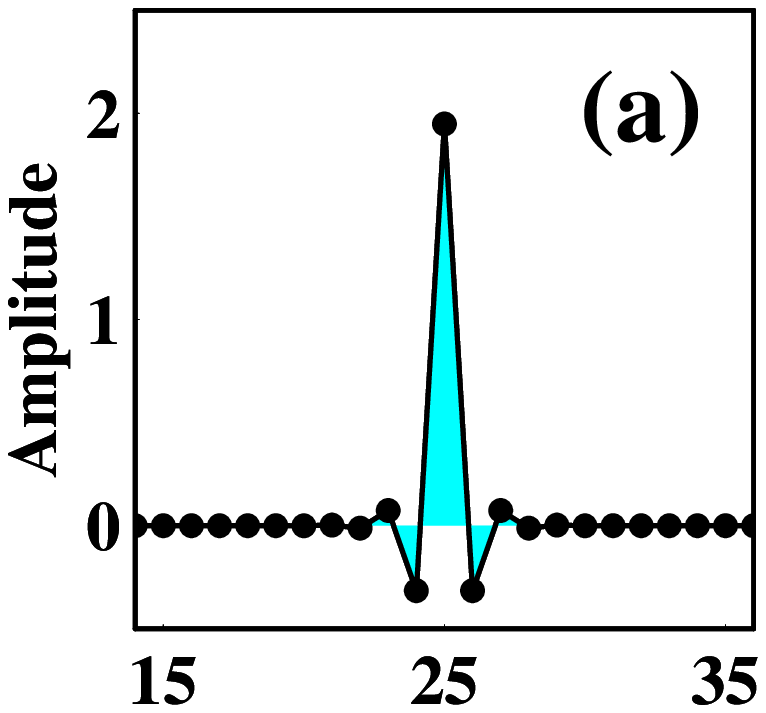}
\includegraphics[scale=0.4,angle=0]{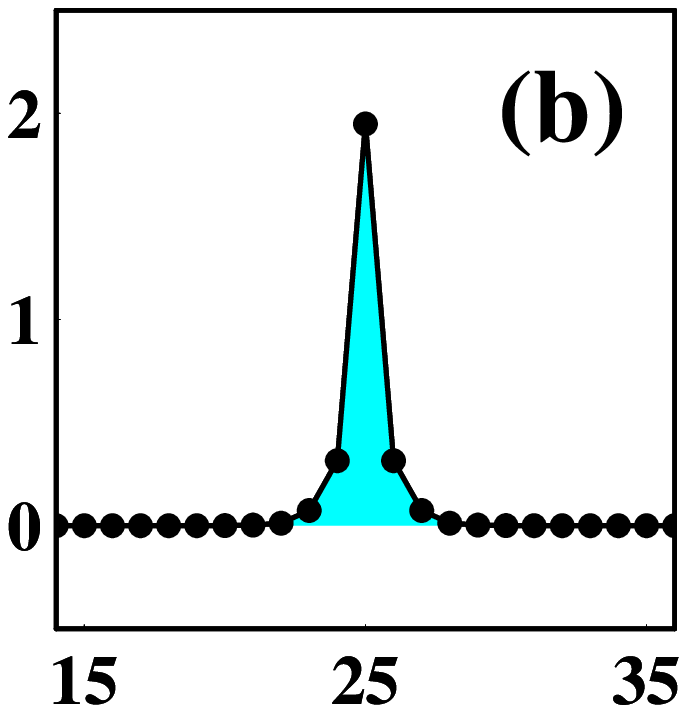}
\end{center}
\caption{
(Color online)
Typical Hamiltonian bulk symmetric breather profiles in a magnetoinductive 
binary chain for 
$\delta=0.8$, $\chi=-1/6$, $\omega=1.5$, and (a) $\lambda=+0.1$; (b) $\lambda=-0.1$. 
These profiles are also obtained by the RWA.
}
\end{figure}

{\em Bulk breathers.-}
Typical bulk Hamiltonian DB profiles obtained with the RWA method
are shown in Figs. 5 and 6.
In Fig. 5, the two single-site symmetric DBs differ in that 
the first one (Fig. 5(a)) is staggered, while the other one  (Fig. 5(b))
is unstaggered. The staggered/unstaggered character of those Hamiltonian 
DBs depends on the sign of the product of the coupling parameter 
and the nonlinearity parameter, $\sigma= sgn(\alpha \lambda)$.  
For $\sigma >0$, that implies either $\alpha=+1$ and $\lambda >0$
or $\alpha=-1$ and $\lambda <0$, the excited DBs are unstaggered.
On the other hand,
for $\sigma <0$, that implies either $\alpha=+1$ and $\lambda <0$
or $\alpha=-1$ and $\lambda >0$, the excited DBs are staggered.
In that figure the DB frequency $\Omega_B =2\pi/T_B$, with $T_B$
the DB period, was chosen to ensure that the DB amplitude at all sites 
is nonzero in the anticontinuous limit.
In Fig. 6, that shows two-site antisymmetric bulk DBs,
the nonlinearity parameter is negative ($\alpha=-1$) implying that
in the anticontinuous limit, only the even sites of the array can have 
a nonzero amplitude.
Again, with appropriate choice of the initial conditions and by changing 
the sign of the coupling parameter $\lambda$ and/or the nonlinearity
parameter $\alpha$ we may also construct multi-site bulk DBs with different
symmetry. It should be also possible to generate a large variety of surface
and bulk Hamiltonian DBs in two dimensional binary arrays, 
just like in planar arrays of identical SRRs \cite{Eleftheriou2008}.
Increased dimensionality offers more possibilities for generating different 
DB types.
\begin{figure}[t!]
\begin{center}
\includegraphics[scale=0.4,angle=0]{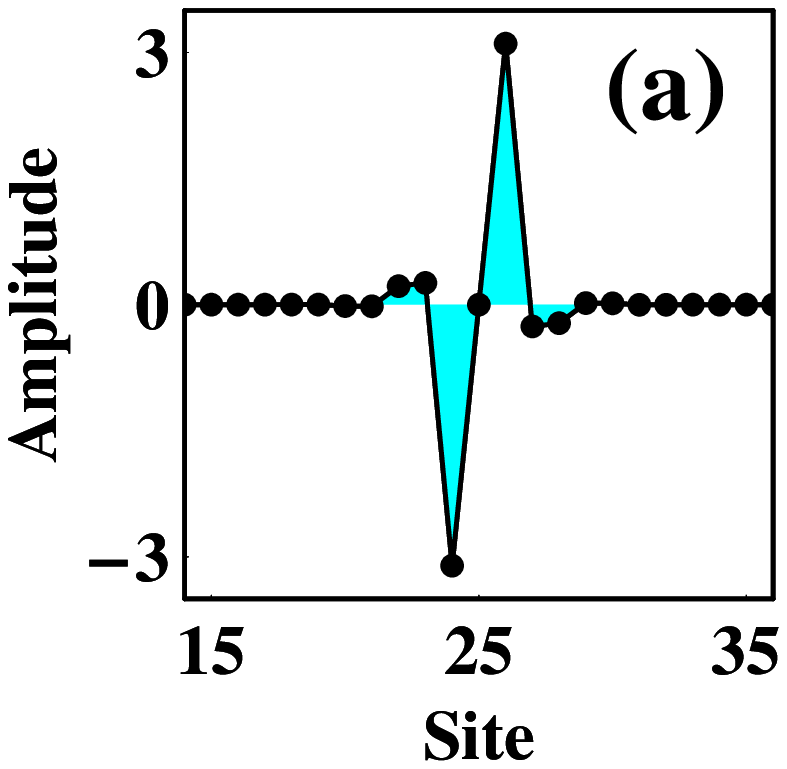}
\includegraphics[scale=0.4,angle=0]{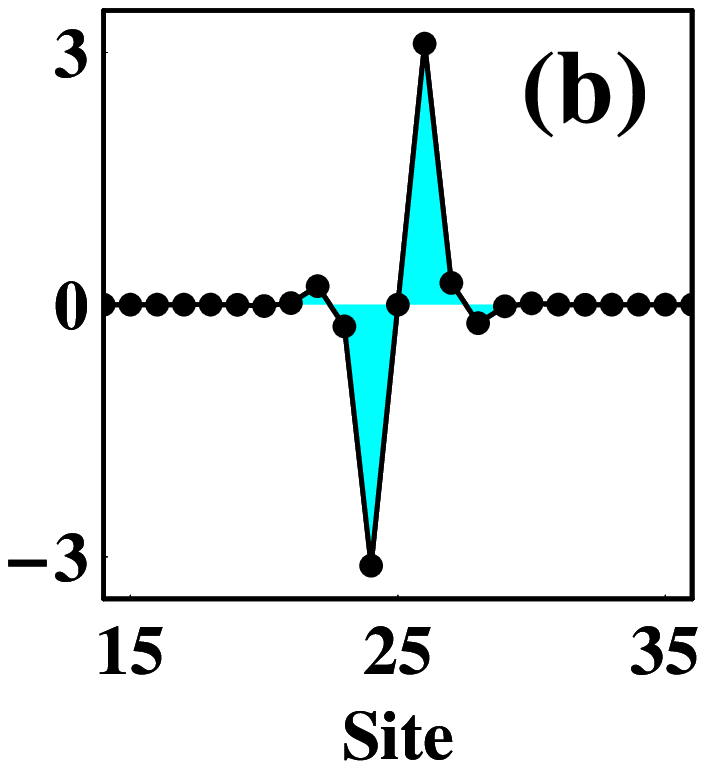}
\end{center}
\caption{
(Color online)
Typical Hamiltonian bulk antisymmetric breather profiles in a magnetoinductive 
binary chain for 
$\delta=0.5$, $\chi=-1/6$, $\omega=0.8$, and (a) $\lambda=-0.2$; (b) $\lambda=+0.2$. 
These profiles are also obtained by the RWA.
}
\end{figure}

\section{Dissipative Breathers in the Rotating Wave approximation
}
We consider the more realistic case of dissipative discrete breathers (DDBs), that  can be excited 
either on the `surface' or the bulk of a 1D binary MM.
In typical experiments that involve MMs, the metamaterial is driven by
an applied electromagnetic field of appropriate polarization. 
In 1D there are two possible geometries for the arrangement of the SRRs 
\cite{Eleftheriou2008};
the planar geometry (as shown in Fig. 1), 
for which the coupling parameter is negative, and the axial geometry
for which the coupling parameter is positive.
That polarization can be chosen such that, for example, the magnetic component
of the field is perpendicular to the SRR planes, while the electric component
is parallel to the sides of the SRRs that do not have a slit.
This choice simplifies physically the situation, since only the magnetic field
excites an emf in the SRRs. Thus, in the equivalent circuit picture,
a binary SRR-based MM in an electromagnetic field can be described by an array
of nonlinear RLC circuits driven by an alternating emf that are coupled through
their mutual inductances.
The losses of the SRRs can be described in this picture 
by an equivalent resistance.
That effective resistance $R$ may actually describe both Ohmic losses of the
SRR as well as radiative losses, if these are relatively low \cite{Kourakis}.
Under those assumptions, the dynamics of the charges $q_n (\tau)$, $n=1,2,....,N$,
is given by Eq. (\ref{12}). In the framework of the RWA method, 
we look for stationary solutions of that equation  in the form 
$q_n(\tau)=q_n \sin(\Omega \tau+\phi_n)$.
By direct substitution of $q_n(\tau)$ into Eq. (\ref{12}) and by making the 
RWA replacement $\sin(\Omega \tau+\phi_{n})^3 \approx (3/4)\ \sin(\Omega \tau+\phi_n)$,
where $q_n$ thereafter denotes the time-independent DB amplitude at the $n$th site and 
$\phi_n$ its phase,
we find that the DB amplitudes and  phases at each site $n$ satisfy the relations
\begin{eqnarray}
\label{21}
 \left[-\Omega^2 (\lambda q_{n+1} + q_n + \lambda q_{n-1}) + \omega_n^2 q_n 
 -\frac{3}{4}\chi \omega_n^6 q_n^3 \right]^2 
 \nonumber \\   
 + \gamma^2\ \Omega^2\ q_{n}^2 = \epsilon_{0}^2 \hspace{1cm} \\
\label{22}
   \phi_n = \nonumber \\
   \tan^{-1}\left[{-\gamma\ \Omega\ q_n
     \over{\omega_{n}^{2}\ q_{n} -\Omega^2 (\lambda q_{n+1} + q_n +\lambda q_{n-1}) 
      -\frac{3\chi}{4} \omega_{n}^{6}\ q_n^3}} \right] \hspace{1cm}
\end{eqnarray}
where $n=1,2,\hdots,N$ and $q_0=q_{N+1}=0$.
The inclusion of dissipation and external driving alters significatively the possible 
DB modes that the binary SRR system can support.
The dissipative DBs possess the character of an attractor for initial conditions
in the corresponding basin of attraction, and they may appear as a result of 
power balance between the incoming power and the intrinsic power loss
\cite{Marin2001,Zueco}
Dissipative DB excitations in SRR-based MMs are of great importance since they alter
locally the magnetic response of the system from diamagnetic to paramagnetic
or vice versa \cite{Lazarides2006,Eleftheriou2008,Eleftheriou2009}.

In the  anticontinuous limit Eqs. (\ref{21}) and (\ref{22}) become
\begin{eqnarray}
\label{23}
 P(q_n) \equiv q_n^2 \left\{ \left[ \omega_n^2 -\Omega^2
 -\frac{3}{4}\chi \omega_n^6 q_n^2 \right]^2 
 + \gamma^2\ \Omega^2\ \right\} = \varepsilon_{0}^2\ \ \   \\
\label{24}
   \phi_n = 
   \tan^{-1}\left[{-\gamma\ \Omega\ 
     \over{ (\omega_n^2  -\Omega^2 ) 
      -\frac{3}{4} \chi \omega_n^6 q_n^2}} \right]\ \ \ 
\end{eqnarray}
where we kept the subscript $n$ to distinguish between oscillators located
either at an odd-numbered site ($n=$odd integer) or even-numbered site
($n=$even integer).
The polynomial $P(q_n)$ is cubic in $q_n^2$
for  general values of the parameters $\delta$, $\chi$, $\Omega$ and $\gamma$,
with  $P(0)=0$.
\begin{figure}[t!]
\noindent
\includegraphics[scale=.55,angle=0]{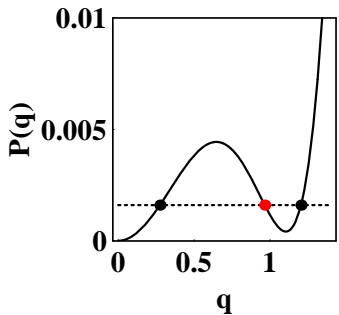}
\caption{
The intersection of $P(q_n)$ (solid line) and $\varepsilon_{0}^2$ (dashed line)
 provides the nonlinear attractor(s) for a single SRR.
Case displayed here corresponds to 
$\delta=1, \chi=1/6, \gamma=0.02, \Omega=0.92, \epsilon_{0}=0.04$. The
black (red) circle represents a stable (unstable) attractor.
}
\end{figure}
Thus, there can be at most three real roots that correspond to attractors
of the SRR oscillator (see Fig. 7), 
from which two are stable (unstable) and one is unstable (stable).
However, by varying a parameter in that four-dimensional parameter space,
two of these solutions may disappear through a pitchfork bifurcation,
leaving behind only one single attractor.
The boundary in parameter space that separates those two cases can be found implicitly
by computing the values of $q_n$ denoted by $q_n^{*}$, for which $dP(q_n) /d q_n =0$
(i.e., the values of $q_n$ that correspond to the local extrema of $P(q_n)$).
Obviously, if $q_n=q_n^{*}$ corresponds to a local minimum (maximum) then 
for $P(q_n^{*}) > \epsilon_0^2$ ($P(q_n^{*}) < \epsilon_0^2$) there is only
one attractor left. Thus, the earlier inequalities determine different regions
in the parameter space where, depending on the values of the parameters, 
we may have either one or three attractors. Thus, we may distinguish
on such a diagram two different `phases' that correspond to either three or one
real solutions. A typical example is shown in the reduced $\epsilon_0 - \Omega$
parameter space in Fig. 8,  
for $\delta=1$ (no RFM) and opposite values of $\chi$,  
while the values of the driving field strength $\epsilon_0$ and the driving frequency 
$\Omega$ are varying. 
There we see clearly the areas where there are either three (inside colored area) or one (outside colored area) attractor(s).
Another example, where the RFM changes from $\delta=0.5$ to $1.5$ is shown 
in Fig. 9 for $\gamma=0.1$ and positive $\chi$. There we observe that the
colored area, corresponding to three attractors,
expands with increasing $\delta$.
\begin{figure}[h]
\noindent
\includegraphics[scale=0.45,angle=0]{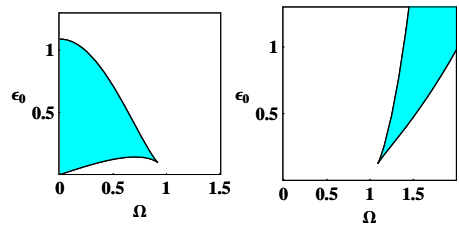}
\caption{
(Color online)
`Phase diagram' in the reduced parameter space $\varepsilon_0 - \Omega$
for a single driven-damped SRR oscillator showing the regions with 
different number of attractors, for $\delta=1$, $\gamma=0.1$, and 
$\chi=+1/6$ (left); $\chi=-1/6$ (right).
Inside (outside) the colored region we have three (one) attractors. 
}
\end{figure}
\begin{figure}[h]
\noindent
\includegraphics[scale=0.3,angle=0]{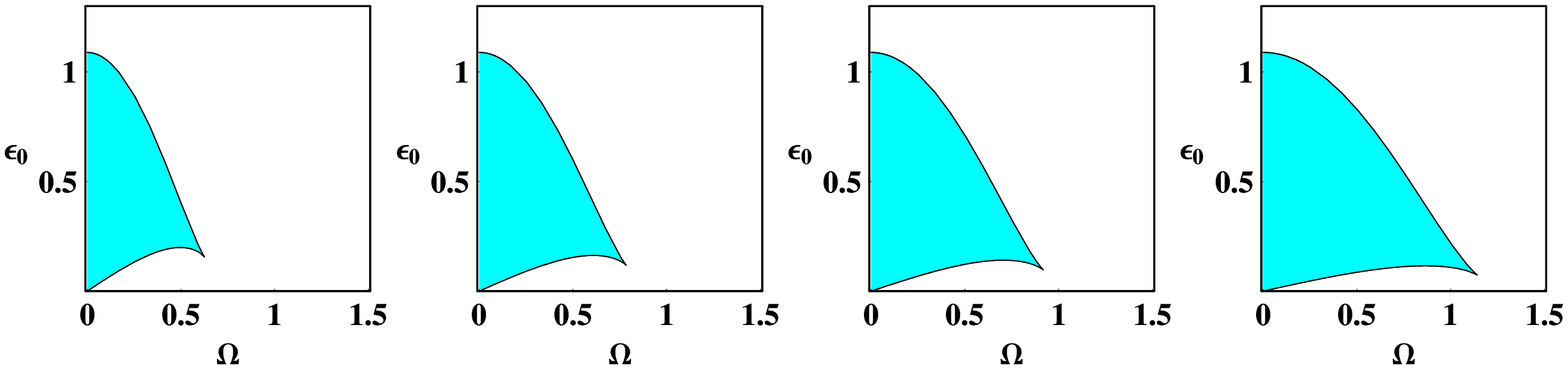}
\caption{
(Color online)
Same as in Fig. 8 for a single driven-damped SRR oscillator
with $\gamma=0.1$, $\chi=+1/6$, and (from left to right)
$\delta=0.5$; $\delta=0.75$; $\delta=1$; $\delta=1.5$.
}
\end{figure}
The values for the stable attractors predicted by the RWA are in excellent 
agreement with those obtained through  dynamical evolution of the charge in
a single SRR oscillator.
For instance, for the parameter set 
$\delta=0.8$, $\Omega=0.92$, $\gamma=0.01$, $\varepsilon_0=0.04$, $\chi=+1/6$, 
and for an even-numbered site $n$, Eq.(\ref{23}) predicts a single attractor at 
$q_1^e=\pm 0.582163$, while for an odd-numbered site it predicts three attractors
at $q_1^o =\pm 1.23531$, $q_2^o =\pm 1.33055$ and $q_3^o=\pm 0.0996811$, 
of which the latter two are stable.
The unstable attractor at $q_1^o$ is not reachable through simple numerical 
integration of the dynamical equation. For the stable attractors, we have checked
with direct numerical integration that their values are practically the same
with those obtained from the RWA approach. 
In general, the presence of dissipation and driving severely limit the possible spatial 
profile of the breathers. The structures tend now to be either strongly localized ones, 
or rather extended, like domain walls  \cite{Lazarides2006,Shadrivov2006}. 
The situation is similar for DDBs in the bulk (Figs.10 and 11). 
As soon as $\delta$ deviates from unity, that is, when we are dealing with a {\em bona fide}
binary chain, the area
in phase space with two stable attractors reduce (increase) as $\delta$ decreases (increases) 
from unity, if the site chosen is an even-numbered one, as can be seen in Fig. 8. 
The opposite behavior occurs for an odd-numbered site.

In order to illustrate how the RWA method works in this case,
we calculate some of typical surface DDBs.
The calculation of bulk DDBs proceeds in the same way,
by simply choosing the central site of the corresponding trivial 
DB that is located at $n=n_B$ somewhere in the bulk.
For a given value of the RFM, we first determine first the attractors
available for each single SRR oscillator, which is located either
at odd- or even-numbered site.
Then, we set up a trivial surface DB which is subsequently continued 
for finite values of $\lambda$. The continuation procedure proceeds in 
exactly the same way as that for Hamiltonian DBs, except that the 
relevant equations are now Eqs. (\ref{21})  and (\ref{22}).
Thus, we can obtain several types of surface DDBs for an interval 
of $\lambda$ up to a maximum, i.e., up to $\lambda=\lambda_{max}$.
For example, for the parameter set 
$\chi=+1/6, \gamma=0.01, \Omega=0.5$, $\delta=2$, and $\varepsilon_0 =0.04$, 
we have stable attractors $q_1^o =\pm 4.06719$ and $q_2^o =\pm 0.160225$ at 
odd-numbered sites,
and $q_1^e =\pm 1.334$ and $q=\pm 0.0228639$ at even-numbered sites.
We can set up a trivial surface DB localized at $n_B =1$ as $q_{1}=4.06719$, 
$q_{2 n}=0.0228639$ and $q_{2n-1}=0.160225$ ($n>1$).
For a trivial surface DB localized at $n_B =2$ we may choose
$q_{2}=1.334$. $q_{2n-1}=-0.160225$ and $q_{2n}=-0.0228639$ ($n>1$). 
Or, for  a trivial surface DB localized at $n_B =3$ we may choose
$q_{2n}=-0.160225$, $q_{2n-1}=-0.160225$ ($n \ne 2$), and $q_{3}=4.06719$.
Continuation of those trivial DDBs up to $\lambda =0.025$ gives the 
surface DDB profiles shown in Fig. 10. 
Of course there also other trivial DDB profiles that we could choose.
Similar bulk DDBs can be obtained from the trivial DDBs given above
only by changing $n_B$ to a value relatively far from the end-points. 
An illustrative example of a bulk DDB localized at an odd-numbered site
is shown in the left panel of Fig. 11, while in the middle and right panels
of Fig. 11 are shown two multi-site DDBs also localized at odd-numbered sites. 
The latter two DDBs have been obtained with appropriate choice of 
a trivial DDB profile.
\begin{figure}[t]
\noindent
\includegraphics[scale=0.4,angle=0]{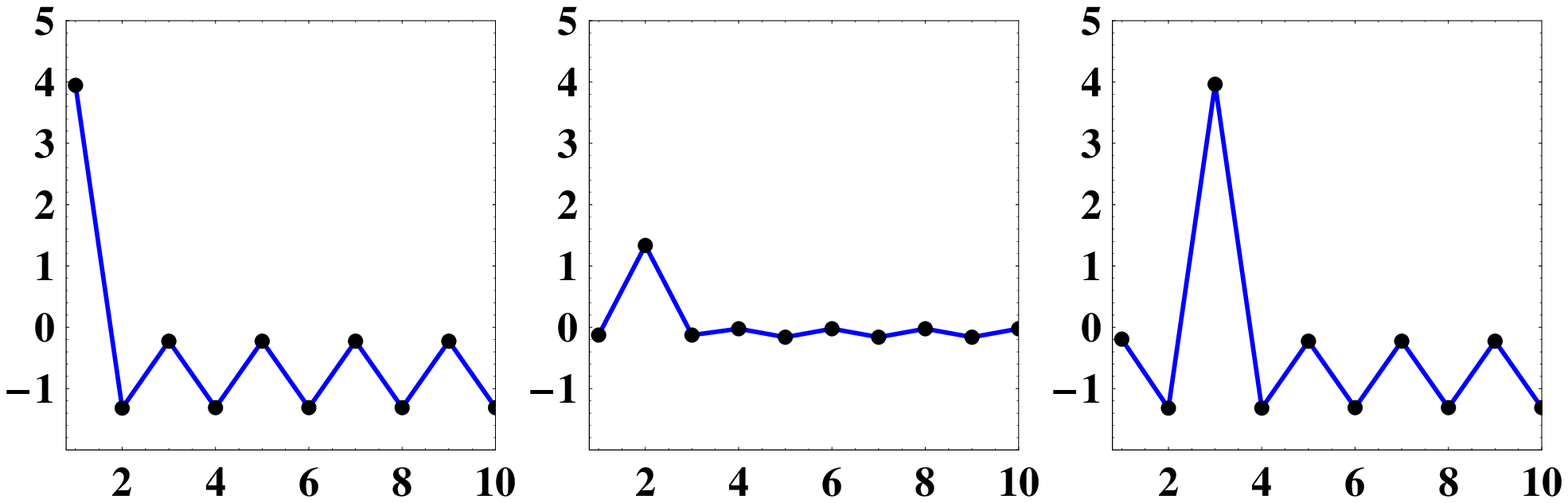}
\caption{
(Color online)
Dissipative surface breather profiles for  $\delta=2$,
$\Omega=0.5$, $\lambda=0.025$, $\chi=+1/6i$, $\gamma=0.01$, 
$\epsilon_{0}=0.04$, that are localized at $n=1$ (left panel);
$n=2$ (middle panel); $n=3$ (right panel).
}
\end{figure}
\begin{figure}[h]
\noindent
\includegraphics[scale=0.4,angle=0]{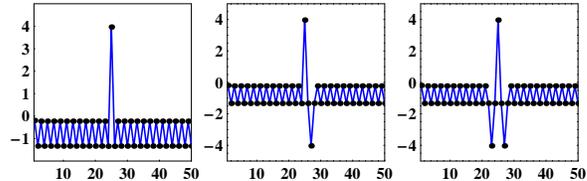}
\caption{
(Color online)
Dissipative bulk breather profiles for  $\delta=2$,
$\Omega=0.5$, $\lambda=0.025$, $\chi=+1/6$, $\gamma=0.01$,
$\epsilon_{0}=0.04$, that are localized at $n=1$ (left panel);
$n=2$ (middle panel); $n=3$ (right panel).
}
\end{figure}

\section{Numerically exact calculations
}
In this Section we construct several types of both energy-conserved
and dissipative DBs which are localized either in the bulk or at the 
surface, using standard numerical algorithms \cite{Marin1996,Aubry1997}. 
Moreover, in the case of a dissipative binary MM we also generate  
DDB excitations through a procedure that can be used for parameter values
where the homogeneous solution is modulationally unstable \cite{Sato2003}. 
In other words, we exploit MI to initiate spontaneous localization of energy
in the binary array \cite{Dauxois}.

{\em Hamiltonian breathers.-}
For the Hamiltonian binary MM, DBs can be constructed from the 
anticontinuous limit of Eqs. (\ref{12}) with $\epsilon_0 =0$ and $\gamma=0$
where all the SRRs are decoupled \cite{Lazarides2006,Eleftheriou2008}.
Using Newton's method we have constructed several types of Hamiltonian,
numerically exact DBs for the 1D binary MM, for different parameter sets.
The obtained Hamiltonian DB profiles are in excellent agreement with 
those obtained with the RWA method.  

{\em Dissipative Breathers.-}
In order to generate DDBs we start from the anticontinuous limit of 
Eq. (\ref{12}), where dissipation and driving are included.
We identify stable attractors of each SRR oscillator, that is 
either located at odd- or even-numbered sites.
For constructing a trivial DDB profile we need to find,
for at least one of the two types of oscillators, two different
amplitude stable attractors. 
For example, for the parameter set
$\delta=2.0$, $\Omega=0.5$, $\gamma=0.01$, $\varepsilon_0=0.04$, $\chi=+1/6$, 
we obtain stable attractors $q_1^o =\pm 4.06719$ and $q_2^o =\pm 0.160225$
at odd-numbered sites,
and $q_1^e =\pm 1.334$ and $q^e=\pm 0.0228639$ at even-numbered sites.
Those values are practically the same with those obtained with the RWA method.
We set up a trivial surface DB localized at $n_B =1$ as $q_{1}=4.06719$, 
$q_{2 n}=0.0228639$ and $q_{2n-1}=0.160225$ ($n>1$).
For  a trivial surface DB localized at $n_B =3$ we may choose
$q_{3}=4.06719$, $q_{2n}=0.0228639$, and $q_{2n-1}=-0.160225$ ($n \ne 2$).
Continuation of those trivial DDBs gives surface DDB profiles up to
$\lambda_{max} \simeq 0.19$. Typical profiles for several values of $\lambda$,
both for DDBs localized at $n=1$ and $n=3$, are shown in Fig. 12.

\begin{figure}[t!]
\noindent
\includegraphics[scale=0.32,angle=0]{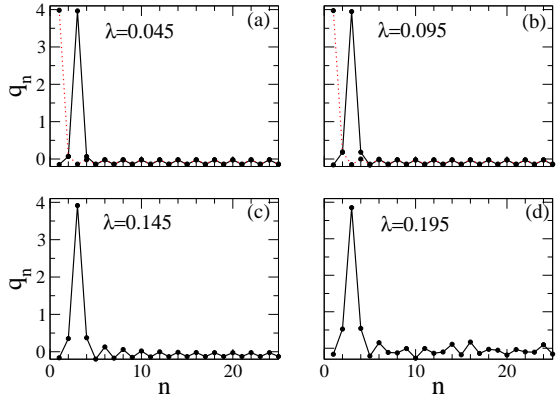}
\caption{
Dissipative breather profiles at maximum amplitude for several 
values of the coupling parameter as shown on the figure which 
are localized at $n=1$ and $n=3$. The parameters are $\Omega=0.5$, $\chi=+1/6$, $\gamma=0.01$, 
$\epsilon_{0}=0.04$, and $\delta=2$.
}
\end{figure}
Another example is given for the parameter set 
$\delta=0.8$, $\Omega=0.92$, $\gamma=0.01$, $\varepsilon_0=0.04$, $\chi=+1/6$, 
where the RWA method predicts a single attractor at $q_1^e=\pm 0.582163$
at even-numbered site oscillators, 
while it predicts three attractors at $q_1^o =\pm 1.23531$, $q_2^o =\pm 1.33055$
and $q_3^o=\pm 0.0996811$, of which the latter two are stable,
for an odd-numbered site oscillators.
These values are also in agreement with those obtained by direct integration
of the single SRR oscillators.
We set up a trivial surface DDB localized at $n_B =1$ as $q_{1}=1.33055$,
$q_{2 n}=0.582163$ and $q_{2n-1}=0.0996811$ ($n>1$), and continue it 
up to $\lambda_{max} \sim 0.07$ where DDBs cease to exist. 
Typical DDB profiles are shown in Fig. 13 for several values of $\lambda$
shown on the figure. A profile for $\lambda =0.072$ which is greater that 
$\lambda_{max}$, where the homogeneous solution is restored, is also 
shown in Fig. 13d.  
The frequency of the DDBs shown here is the same with that of the driver,
i.e., $\Omega_B \equiv 2\pi/T_B = \Omega$.
However, the phase differences of the SRR oscillators in the array with respect
to the driving field are generally different for each oscillator, 
as can be observed in Fig. 14, where the time-evolution of $q_1 - q_4$ is
followed for approximately two periods $T_B$ of the DDB oscillation.
Note also that the time-evolution seems practically sinusoidal (harmonic),
that may not be necessarily true for DDBs obtained with some other parameter set.
\begin{figure}[h!]
\includegraphics[scale=0.32,angle=0]{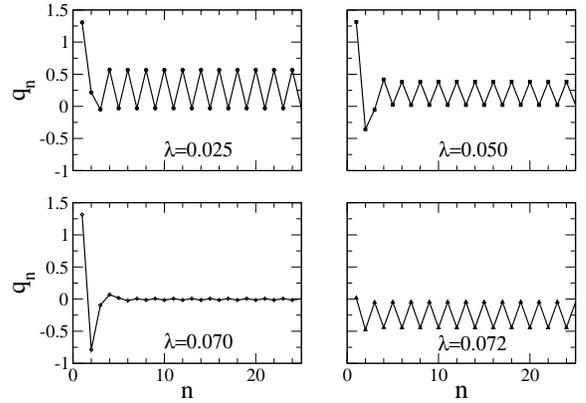}
\caption{
Dissipative breather profiles at maximum amplitude for several 
values of the coupling parameter as shown on the figure which 
are localized at the surface (at $n=1$), along with an almost 
uniform solution for $\lambda=0.072$ just above the value of
$\lambda_{max}$ for this particular parameter set.
The parameters are $\Omega=0.92$, $\chi=+1/6$, $\gamma=0.01$, 
$\epsilon_{0}=0.04$, and $\delta=0.8$.
}
\end{figure}

\begin{figure}[h!]
\includegraphics[scale=0.45,angle=0]{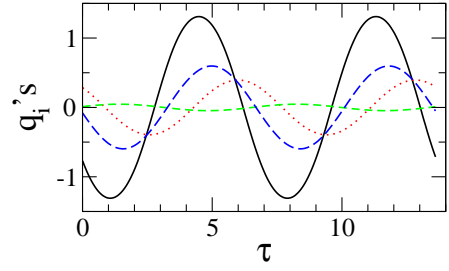}
\caption{
(Color online)
Time-dependence of $q_1$ (black solid curve), $q_2$ (red dotted curves),
$q_3$ (green short-dashed curve), and $q_4$ (blue dashed curves),
for a surface breather localized at $n=1$ and  $\lambda=0.03$.
The other parameters are the same with those in the caption of Fig. 14.
}
\end{figure}

{\em Dissipative Breathers by frequency chirping.-}
For a frequency gapped linear spectrum, some of the linear modes become
unstable at large amplitude. If the curvature of the dispersion curve
in the region of that mode is negative and the lattice potential is hard
then, the large amplitude mode becomes unstable with respect to formation
of a DB in the gap above the linear spectrum \cite{Sato2003,Sato2006}.
Below we exploit MI in order to generate spontaneously DDBs in the binary array.
The procedure that is followed is shortly described below
\cite{Sato2003,Sato2006}.
For the parameters in the captions of Fig. 15 and 16, the top of the upper linear
band is located at $\Omega \simeq 1.42$ where the curvature is negative.
Moreover, the SRRs are subjected to on-site potentials that are hard
(for $\chi <0$). The (large amplitude) driver is initiated with its
frequency just below $\Omega$ and is then chirped with time to produce
enough vibrational amplitude to induce MI of the uniform mode, which
then leads to spontaneous DDB generation.
At the end of the frequency chirping phase, the driver frequency is
well above $\Omega$, and only supplies energy into the formed DDB(s).
During that phase, a large number of DDBs may be generated, which can move
and collide and eventually coalesce into a small number of high amplitude
DDBs that are frequency locked to the driver and, because of that,
they are trapped at particular SRRs.
After that, the driver frequency is kept constant and the high amplitude
DDBs (and even some low-amplitude ones) continue to receive energy
falling into a stationary state. When the driver is switched off
all DDBs die out in a short time interval.

\begin{figure}[t!]
\includegraphics[scale=0.65,angle=0]{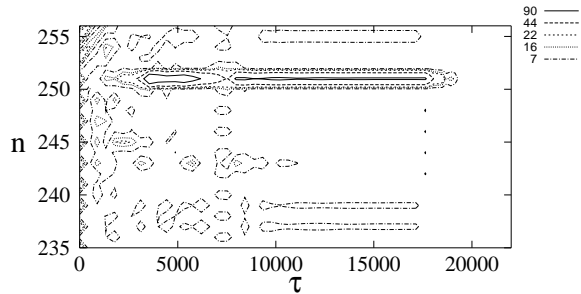}
\caption{
Contours of the energy density ${\cal H}_n$ on the $\tau - n$ plane
for a strongly driven binary magnetic metamaterial, with $\delta=2$,
$\Omega=1.42$, $\chi=-1/6$, $\gamma=0.001$, $\lambda=0.05$,
and $\varepsilon_{0}=3.0$. 
}
\end{figure}
In Figs. 15 and 16, the contours of the energy density ${\cal H}_n$ on the $\tau -n$
plane identify the evolution of the DDBs formed by that procedure.
The chirping phase lasts for $2000~T_0$ time units
($T_0 =2\pi / \Omega$), where the frequency varies linearly from
$\Omega_i = 0.997 \Omega$ to $\Omega_f = 1.020 \Omega$.
The driver is subsequently kept at constant frequency $\Omega_f$ until
it is switched off after another $2000~T_0$ time units.
Figs. 15 and 16 correspond to the regions of the binary MMs where several
DDBs have survived after the chirping phase. In Fig. 15 we clearly observe 
a high amplitude DDB at $n=251$, along with some other DDBs
of considerably lower amplitude, that survive until the end of the
constant frequency phase. In Fig. 16, where the binary MM is driven not 
as strongly as that in Fig. 15, we observe one relatively low-amplitude DDB
which however survives until the end of the constant frequency phase.
  It is possible that the procedure described above, which relies on the
MI of the large amplitude linear modes, can be used for the generation of DDBs
in other magnetoinductive systems as well \cite{Lazarides2008b,Tsironis2009},
in their binary versions.

\section{Conclusion}
We presented detailed analysis for induced nonlinear localization in binary 
nonlinear magnetic metamaterials.
The systems we analyzed are one dimensional and consist of two types of SRRs; 
this configuration leads to a
linearized  magnetoinductive system with two optical bands separated by a gap.  
When nonlinearity is also taken into account nonlinear localized modes of 
discrete breather type may be generated in the gaps.  As in the case
where all units are identical~\cite{Lazarides2006,Eleftheriou2008} depending 
on the 
\begin{figure}[t!]
\includegraphics[scale=0.65,angle=0]{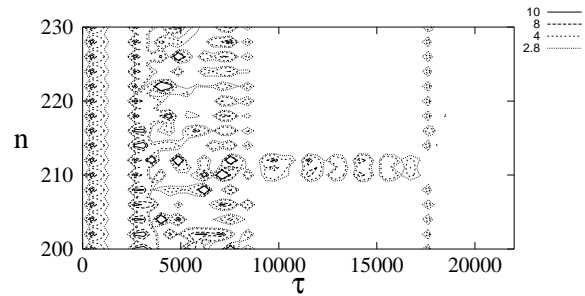}
\caption{
Contours of the energy density ${\cal H}_n$ on the $\tau - n$ plane
for a strongly driven binary magnetic metamaterial, with $\delta=2$,
$\Omega=1.42$, $\chi=-1/6$, $\gamma=0.001$, $\lambda=0.05$, 
and $\varepsilon_{0}=2.0$. 
}
\end{figure}
parameter regime 
these breathers may be left-handed in a right-handed background or the opposite.  
We focussed on both the Hamiltonian as well as the dissipative case; 
the latter is clearly the most interesting physically since it corresponds 
to true propagation of waves in the medium.  
We used two approaches, one based on the rotating wave approximation while 
the second on exact numerics using the
breather analysis from the anticontinuous limit.  
The comparison of the two showed that the RWA is in most cases
a good approximation for a relatively accurate breather construction.

In the Hamiltonian case we found two types of breathers with even or odd local 
symmetry depending on the sign of the
product of the coupling parameter and the nonlinearity parameter.  
Both types may exist in the bulk but also in the boundary of the chain; 
the latter form surface breathers.  A similar situation occurs also in the 
dissipative case where depending on the number of attractors of the single 
driven nonlinear oscillator system we have different type of dissipative breathers.
Both bulk and surface breathers appear with corresponding symmetries.  

The binary structure of the lattice allows for generation of breathers through 
direct external induction. 
This is accomplished through frequency chirping to the desired frequency.  
In the process of frequency modulation induction of plane wave instability 
occurs that leads to breather generation.  
These breathers move around in the lattice, collide, some decay and eventually 
a single ``large" breather is left in the metamaterial.  
This method of breather generation is direct and may be used for experimental 
breather investigation in metamaterials. 

\section*{Acknowledgments}
One of us (MIM) acknowledges support from Fondecyt Grant 1080374 of Chile.


\begin{thebibliography}{99}
\bibitem{ST}
A. J. Sievers and S. Takeno, Phys. rev. Lett. {\bf 61}, 970 (1988).

\bibitem{Flach}
For a recent review see
S. Flach, A.V. Gorbach, Phys. Rep. {\bf 467}, 1 (2008).

\bibitem{MacKay}
  R. S. MacKay and S. Aubry, Nonlinearity {\bf 7}, 1623 (1994).

\bibitem{Aubry1997}
  S. Aubry, Physica D {\bf 103}, 201 (1997).

\bibitem{Marin1996}
  J. L. Mar\'{\i}n and S. Aubry, Nonlinearity {\bf 9}, 1501 (1996). 

\bibitem{Marin2001}
  J. L. Mar\'{\i}n, F. Falo, P. J. Mart\'{\i}nez, and L. M. Flor\'{\i}a, 
  Phys. Rev. E {\bf 63}, 066603 (2001).
 
\bibitem{Zueco}
  D. Zueco, P. J. Mart\'{\i}nez, L. M. Flor\'{\i}a, and F. Falo, 
  Phys. Rev. E {\bf 71}, 036613 (2005).

\bibitem{Eisenberg}
 H. S. Eisenberg, Y. Silberberg, R. Morandotti, A. R. Boyd,
 and J. S. Aitchison,
 Phys. Rev. Lett. {\bf 81}, 3383 (1998).

\bibitem{Swanson}
  B. I. Swanson, J. A. Brozik, S. P. Love, G. F. Strouse, A. P. Shreve,
  A. R. Bishop, W.-Z. Wang, and M. I. Salkola,
  Phys. Rev. Lett. {\bf 82}, 3288 (1999).

\bibitem{Russell}
  F. M. Russell, and J. C. Eilbeck,
  Europhys. Lett. {\bf 78}, 10004 (2007).

\bibitem{Schwarz}
  U. T. Schwarz, L. Q. English, and A. J. Sievers,
  Phys. Rev. Lett. {\bf 83}, 223 (1999).

\bibitem{Trias}
  E. Tr\'ias, J. J. Mazo, and T. P. Orlando,
    Phys. Rev. Lett. {\bf 84}, 741 (2000).

\bibitem{Sato2003}
  M. Sato, B. E. Hubbard, A. J. Sievers, B. Ilic, D. A. Czaplewski, and
  H. G. Graighead,
  Phys. Rev. Lett. {\bf 90}, 044102 (2003).

\bibitem{Sato2007}  
  M. Sato and A. J. Sievers,
  Phys. Rev. Lett. {\bf 98}, 214101 (2007).

\bibitem{TA}
G. P. Tsironis and S. Aubry, Phys. Rev. Lett. {\bf 77}, 5225 (1996).

\bibitem{R}
K. O . Rasmussen, S. Aubry, A. R. Bishop and  G. P. Tsironis,
Eur. Jour. Phys. B {\bf 15}, 169 (2000).

\bibitem{Dauxois}
  T. Dauxois and M. Peyrard,
  Phys. Rev. Lett. {\bf 70}, 3935 (1993).

\bibitem{Hennig2007}
  D. Hennig, L. Schimansky-Geier, and P. H\"anggi,
  Europhys. Lett. {\bf 78}, 20002 (2007).
  
\bibitem{Hennig2008}
   D. Hennig, S. Fugmann, L. Schimansky-Geier, and P. H\"anggi,
   Acta Physica Polonica B {\bf 39}, 1125 (2008).   

\bibitem{Sato2006}
  M. Sato, B.E. Hubbard, A.J. Sievers,
    Rev. Mod. Phys. 78 (2006) 137.

\bibitem{Gorkunov2002}
  M. Gorkunov, M. Lapine, E. Shamonina, and K. H. Ringhofer,
  Eur. Phys. J. B {\bf 28}, 263 (2002).

\bibitem{Linden}
  S. Linden, C. Enkrich, G. Dolling, M. W. Klein, J. Zhou,
  T. Koschny, C. M. Soukoulis, S. Burger, F. Schmidt, and M. Wegener,
  IEEE J. Selec. Top. Quant. Electron. {\bf 12}, 1097 (2006).

\bibitem{Yen}
  T. J. Yen, W. J. Padilla, N. Fang, D. C. Vier, D. R. Smith,
  J. B. Pendry, D. N. Basov, and X. Zhang,
  Science {\bf 303}, 1494 (2004).

\bibitem{Katsarakis}
  N. Katsarakis, G. Konstantinidis, A. Kostopoulos, R. S. Penciu,
  T. F. Gundogdu, M. Kafesaki, E. N. Economou, Th. Koschny,
  and C. M. Soukoulis,
  Opt. Lett. {\bf 30}, 1348 (2005).

\bibitem{Hand}
  T. H. Hand and S. A. Cummer,
  J. Appl. Phys. {\bf 103}, 066105 (2008).

\bibitem{Powell}
  D. A. Powell, I. V. Shadrivov, Yu. S. Kivshar, and M. V. Gorkunov,
  Appl. Phys. Lett. {\bf 91}, 144107 (2007).

\bibitem{Shadrivov2006}
  I. V. Shadrivov, S. K. Morrison, and Yu. S. Kivshar,
  Opt. Express {\bf 14}, 9344 (2006).

\bibitem{Shadrivov2008}
  I.V. Shadrivov, A.B. Kozyrev, D. van der Weide, Yu.S. Kivshar,
  Appl. Phys. Lett. {\bf 93}, 161903 (2008).

\bibitem{Lazarides2006}
  N. Lazarides, M. Eleftheriou, and G. P. Tsironis, 
  Phys. Rev. Lett. {\bf 97}, 157406 (2006). 

\bibitem{Eleftheriou2008}
  M. Eleftheriou, N. Lazarides, and G. P. Tsironis, 
  Phys. Rev. E {\bf 77}, 036608 (2008).

\bibitem{Lazarides2008}
  N. Lazarides, G. P. Tsironis, and Yu. S. Kivshar,
  Phys. Rev. E {\bf 77}, 036608 (2008).

\bibitem{Eleftheriou2009}
  M. Eleftheriou, N. Lazarides, G. P. Tsironis, and Yu. S. Kivshar, 
  arXiv:0903.2130v1 [cond-mat.mtrl-sci] (subm. to Phys. Rev. E).

\bibitem{Shadrivov2006b}
  I. V. Shadrivov, A. A. Zharov, N. A. Zharova, and Yu. S. Kivshar,
  Photonics Nanostruct. Fundam. Appl. {\bf 4}, 69 (2006).

\bibitem{Kourakis}
  I. Kourakis, N. Lazarides, and G. P. Tsironis,
  Phys. Rev. E {\bf 75}, 067601 (2007).

\bibitem{Molina2006b}
   M. I. Molina, R. A. Vicencio, and Yu. S. Kivshar,
  Opt. Lett. {\bf 31}, 1693 (2006).

\bibitem{Kivshar}
   Yu. S. Kivshar and M. I. Molina,
   Wave Motion {\bf 45}, 59 (2007).

\bibitem{Gorkunov2006}
   M. V. Gorkunov, I. V. Shadrivov, and Yu. S. Kivshar,
   Appl. Phys. Lett. {\bf 88}, 071912 (2006).

\bibitem{Shamonina}
  E. Shamonina, V. A. Kalinin, K. H. Ringhofer, and L. Solymar,
  J. Appl. Phys. {\bf 92}, 6252 (2002).

\bibitem{Shadrivov2007}
  I. V. Shadrivov, A. N. Reznik, and Yu. S. Kivshar,
  Physica B {\bf 394} 180 (2007).

\bibitem{Zharov}
   A. A. Zharov, I. V. Shadrivov, and Y. S. Kivshar,
   Phys. Rev. Lett. {\bf 91}, 037401 (2003).

\bibitem{Molina2006}
  M. I. Molina, I. L. Garanovich, A. A. Sukhorukov, and Yu. S. Kivshar, 
  Opt. Lett. {\bf 31}, 2332 (2006).

\bibitem{Lazarides2008b}
  N. Lazarides, G. P. Tsironis, and M. Eleftheriou,
  Nonlin. Phen. Compl. Syst. {\bf 11}, 250 (2008).

\bibitem{Tsironis2009}
 G. P. Tsironis, N. Lazarides, and M. Eleftheriou,
 PIERS Online {\bf 5}, 26 (2009).

\end{thebibliography}
\end{document}